\documentstyle[preprint,aps,floats,epsf]{revtex}

\newcommand{\be}{\begin{equation}}
\newcommand{\ee}{\end{equation}}
\newcommand{\bea}{\begin{eqnarray}}
\newcommand{\eea}{\end{eqnarray}}

\def\href#1#2{#2}

\def\IZ{\relax\ifmmode\mathchoice
{\hbox{\cmss Z\kern-.4em Z}}{\hbox{\cmss Z\kern-.4em Z}} {\lower.9pt\hbox{\cmsss
Z\kern-.4em Z}} {\lower1.2pt\hbox{\cmsss Z\kern-.4em Z}}\else{\cmss Z\kern-.4em
Z}\fi}
\def\IR{\relax{\rm I\kern-.18em R}}
\font\cmss=cmss10 \font\cmsss=cmss10 at 7pt

\begin{document}

\preprint{SU-ITP-97-43}

\title{On 6d ``Gauge" Theories with Irrational Theta Angle}

\author{\bf Barak Kol}
\address{Department of Physics\\
Stanford University\\ Stanford, CA 94305, USA\\
\tt{barak@leland.stanford.edu}}

\maketitle
\begin{abstract}
A recent proposal for 6d ``gauge" theories with rational theta angle is
discussed. These theories were constructed from $n_C$ coinciding $(p,q)$
5-branes of IIB in the limit of vanishing string constant. They have $(1,1)$
supersymmetry and the low energy theory is 6d Yang-Mills with gauge group
$SU(n_C)$ and a rational theta angle $\theta _6/2\pi=f/q$, where $fp=1$ (mod
$q$). By changing the point of view, considering the $(p,q)$ 5-branes to be
D5-branes, the 6d theta angle is identified with a 10d theta angle. This point
of view together with some assumptions suggests a generalization of the previous
limit to arbitrary theta. This limit seems to define a decoupled 6d theory even
though the 10d theory does not become free in general.
\end{abstract}
\newpage
Six dimensional Theories without gravity were introduced in
\cite{W_com,Strominger}. Related ideas were discussed in studies of black holes,
BPS charges and strings inside 5-branes \cite{DVV1,DVV2,Maldacena}. These
theories were found to play a role in matrix M theory \cite{Rozali,BRS}.
Consider those with (1,1) supersymmetry. Such theories reduce in low energies to
6d Yang-Mills, and their Lagrangian includes the following terms
\be
{\cal L}={1 \over g_6^2}F^2 + {\theta _6 \over 2\pi} F \wedge F \wedge F + \dots
\ee
where F is the field strength, $g_6$ is the 6d coupling constant, and $\theta
_6$ is a 6d theta angle, related to the $\pi_5$ homotopy of the gauge group. The
energy scale in the 5-brane is set by $1/g_6^2$, the tension of the instanton
inside the 5-brane. Some constructions of these theories in terms of the
5-branes of Type IIB \cite{Seiberg,Witten} will be discussed here.

\begin{figure}
\centerline{\epsfxsize=160mm\epsfbox{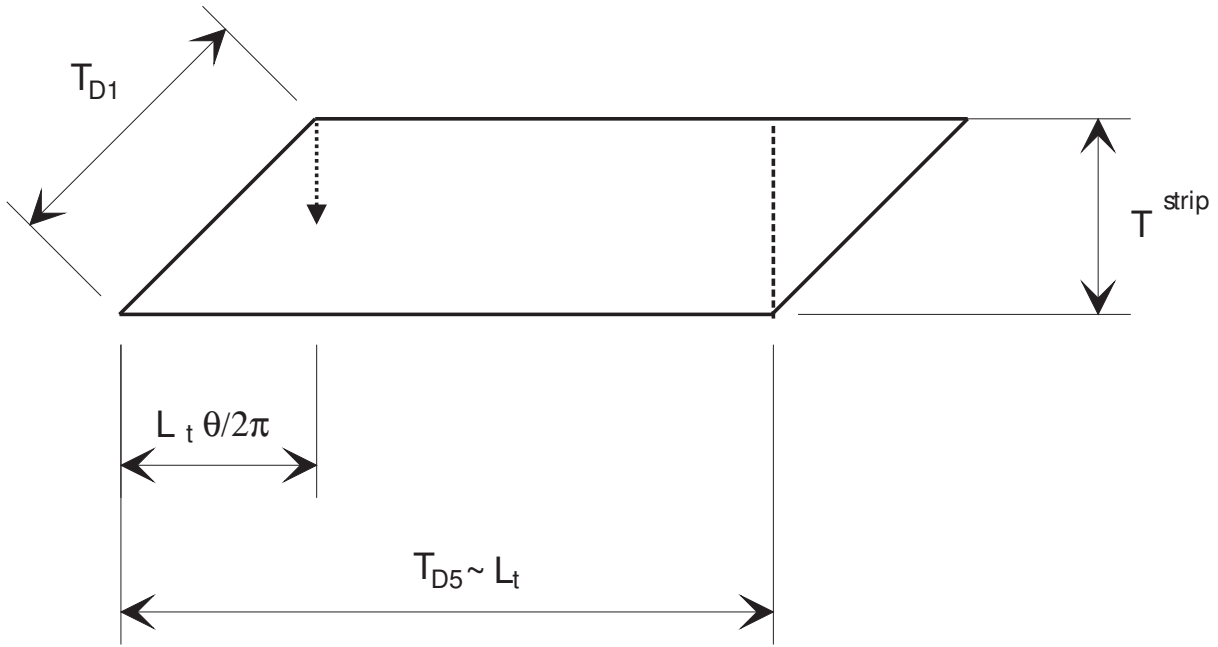}}
\medskip
\caption{The small torus of IIB with base $L_t$ and a theta angle. The shown dimensions are proportional to the D5-brane tension $T_{D5}$, the strip tension $T^{strip}$ (dashed), and the D-string tension $T_{D1}$. The limit $Im(\tau) \to 0$ is demonstrated by the dotted arrow.}
\label{torus}
\end{figure}

Let us set the notations to discuss Type IIB. Denote the complex scalar of IIB
by $\tau={\theta_B \over 2\pi}+i/\lambda$, where $\lambda$ is the string
coupling and $\theta _B$ is the IIB theta angle. Type IIB on a large circle of
size $L_B$ with a complex scalar $\tau$ is the limit of M theory compactified on
a small torus with base $L_t$ and modular parameter $\tau$
\cite{Schwarz,Aspinwall} (figure (\ref{torus})). Let $A_r$ denote the torus area
$A_r=L_t^2Im(\tau)$. The compactification sizes are related through
$2\pi/L_B=A_r T_M=L_t^2Im(\tau)M_{11}^3/(2\pi)^2$, where $T_M$ is the M-membrane
tension
\footnote{The conventions used for $2\pi$ factors are : $T_M=M_{11}^3/(2\pi)^2,
T_{M5}=M_{11}^6/(2\pi)^5,T_{Dp}=1/[\lambda(2\pi)^p \alpha '^{(p+1)/2}]$.} , and
$M_{11}$ is the 11d Planck length. The 10d Planck length $M_p$ is given by $L_B
M_p^8= A_r M_{11}^9$. An expression that depends on the M theory variables
survives the IIB limit if when translated to IIB variables it is independent of
$L_B$. A $(p,q)$ 5-brane
\footnote{Wrapping M-branes around $(p,q)$ cycles produces branes of IIB. A $(1,0)$
string is the fundamental string with tension $T_s=L_t T_M$, whereas a $(0,1)$
string is the D-string. The tension of a general $(p,q)$ string is given by
$T_{p,q}^{string}=\left| p+q\tau \right|L_t T_M$. A general $(p,q)$ 5-brane has
tension $T_{p,q}^{5-brane}=\left| p+q\tau \right|Im(\tau )T_s^3/(2\pi)^2=\left|
p+q\tau \right|{L_t \over L_B} M_{11}^6/(2\pi)^5$, where $(1,0)$ is the D5-brane
and $(0,1)$ is the NS5-brane.} can have an instantonic string inside it. This
instanton was called a ``strip" in \cite{AHK} since its M theory origin is a
membrane stretched over an interval with endpoints on the M5-brane (figure
(\ref{torus})). The strip tension in a general $(p,q)$ 5-brane with a general
$\tau$ is given by
\be
T_{p,q}^{strip}={A_r \over \left| p+\tau q\right|L_t}T_M={Im(\tau) \over \left|
p+\tau q\right| }T_s.
\label{t_strip}
\ee

In \cite{Seiberg} Seiberg introduced a number of new 6d constructions. The
relevant one to us is constructed from
\bea
\text{$n_C$ coinciding NS5-branes} \nonumber \\
\text{taking the limit  } \lambda \to 0.
\label{caseS}
\eea
The low energy gauge group is $SU(n_C)$. For the NS5-brane the tension of the
instanton inside the 5-brane, the only energy scale,  is (\ref{t_strip})
$T_{NS5}^{strip}=1/g_6^2=T_s$. Since this tension is independent of $\lambda$,
it was argued that for $\lambda
\to 0$, the bulk becomes free and a non-trivial 6d theory decouples from it.

To describe this theory in the IIB limit of M theory, we should perform the
limit $\lambda \to 0$ while \underline{fixing the compactification size in 6d
units} $L_B\sqrt{T^{strip}}=\epsilon^{-1}$. For a theory on an NS5-brane the
limit translates to $L_t \to 0$, $\tau=i\epsilon/L^{3/2}(2\pi/T_M^{3/2})$. After
taking this limit we can take the decompactification limit $\epsilon \to 0$.

The appearance of \underline{fractional instantons} of tension $T \sim
T^{strip}/n_C$ \cite{MS} is natural in this picture\footnote{The author thanks
I. Klebanov, J. Maldacena and A. Rajaraman for discussing this point.}. In 11d
the circle transverse to the M5-branes is small and the M5-branes could be
distributed around it in various ways, while still coinciding in the 10d sense.
A general configuration, roughly equally spaced around the circle, results in
$n_C$ fractional strips. These correspond to membranes ending on adjacent
M5-branes. The existence  of fractional strips can be ignored in the following
discussion since $n_C$ will be fixed.

The discussion was generalized by Witten \cite{Witten} to describe theories
parametrized by integers $f,q,n_C$ where $f$ and $q$ are relatively prime and $0
\leq f \leq q-1$. The low energy description of these theories has $SU(n_C)$
gauge group, and a rational theta angle $\theta_6/2\pi=f/q$.  Define integers
$p,e$ such that
\be
M=
\left[\matrix{
p & e \cr q & f \cr }\right]
\in SL(2,\IZ).
\label{matrix}
\ee
Explicitly, $p$ is given by $pf=1$ (mod $q$), and we have the freedom $(p,e) \to
(p,e)+n(q,f), n \in \IZ$. (Note: the notation used here is slightly different
than in \cite{Witten}
). These theories are constructed in type IIB as the theory of
\bea
\text{$n_C$ coinciding $(p,q)$ 5-branes}  \nonumber \\
\text{taking the limit  } \tau \to i\infty,  \nonumber \\
\text{or } L_t \to 0, \tau \sim {i\epsilon \over L^{3/2}\sqrt{q}}\cdot {2\pi \over T_M^{3/2}}.
\label{caseW}
\eea
As in the previous case the bulk theory becomes free, and the 6d is decoupled
and non-trivial.

Let us change the point of view and consider the $n_C$ 5-branes to be $n_C$
D5-branes, through an $SL(2,\IZ)$ transformation. Explicitly, the required
transformation is $M^{-1} (\ref{matrix})$. Then the construction consists of
\bea
&& \text{$n_C$ coinciding D5-branes} \nonumber \\
\text{taking the limit  } && \tau \to 0 \text{ on the imaginary axis, for the first case (\ref{caseS}),} \nonumber \\
\text{and } && \tau \to f/q \text{ with fixed real part, for the second case (\ref{caseW}),} \nonumber \\
\text{or } && L_t \to \infty, \tau = Re(\tau)+i{\epsilon ^2 \over L_t^3}\cdot {(2\pi)^2 \over T_M}.
\eea
The advantage in this point of view is the direct relation
\be
\theta _B=\theta_6
\ee
which can be read off the D5-brane action. Note that in this limit $\lambda \to
\infty$ which is surprising at first.

Let us analyze the conditions for the 10d theory to be free and for the 6d
theory to decouple. \underline{The 10d theory is free} if there is some $(l,k)$
string with a small string coupling, so its tension satisfies
\be
T_{l,k}^{string} \ll M_p^2.
\label{free}
\ee
In the limit $\tau \to f/q$ it is the $(f,-q)$ string that becomes light. Its
tension, as $L_t \to \infty$ is
\be
T_{f,-q}^{string} \sim q Im(\tau) L_t T_M={q\epsilon ^2\over L_t^2} \cdot
{(2\pi)^2\over T_M}.
\label{qtension}
\ee
The tension of the strip, $T^{strip}$ and the tension of the 5-brane, $T_5$, are
related through
\be
T^{strip} \cdot T_5 =  M_p^8.
\ee
This relation is a result of Dirac quantization in 11d between the M-membrane
and its magnetic dual, the M5-brane. So the energy scales in the 6d theory are
characterized by a single scale, and it seems that \underline{the condition for
the 6d theory to decouple} is
\be
T^{strip}=1/g_6^2={A_r \over \left| p+\tau q\right|L_t}T_M \ll M_p^2.
\label{decouple}
\ee

From this point of view \underline{a generalization to irrational theta angles
suggests itself}. Consider taking
\bea
\text{$n_C$ coinciding D5-branes} \nonumber \\
\text{and the limit } \tau \to {\theta_ 6 \over 2\pi} \nonumber \\
\text{or } L_t \to \infty, \tau = {\theta_ 6 \over 2\pi} +i{\epsilon ^2 \over L_t^3}\cdot {(2\pi)^2 \over T_M}.
\eea
$\theta _6$ is arbitrary now and $p,q$ are not required. In order to consider
the conditions (\ref{free},\ref{decouple}) we should measure distances on the
torus plane divided by $M_p^2$. In these units the limit turns out to be $L_t
\to \infty, Im(\tau) = 1/L_t$ (suppressing $M_{11}$), so that the area of the torus is fixed in these units.
We see that the condition for the 6d theory to decouple (\ref{decouple}) is
satisfied. Assuming the mentioned procedures for defining 6d theories are
correct, and that (\ref{decouple}) is indeed the correct condition for a 6d
theory to decouple, we have a theory with an arbitrary theta angle. This became
possible by considering tori that do not necessarily generate a square lattice.

This theory does not allow to define (in general) a limit in which the bulk is
free. It turns out that taking the limit for $\tau$ while measuring tensions in
Planck units corresponds to the torus degenerating while keeping its area fixed.
To have a free limit we should look for light strings. Consider an $(l,k)$
string. Its tension is $\left|(l+k{\theta_ 6
\over 2\pi})L_t+ik/L_t \right|$ in 10d Planck units. As a function of $L_t$ it
has a minimum value of
\be
T_{l,k}=\sqrt{(k(l+k{\theta_ 6 \over 2\pi})} \sqrt{2}.
\ee
An irrational number, $x$, is said to be \underline{approximated by rationals to
order $n$} \cite{Niven} if $\exists c>0$, and infinitely many integers $k$ such
that
\be
\left|kx \text(mod 1) \right|< {c \over k^{(n-1)}}.
\ee
We see that only if the degree of rational approximation for $\theta_ 6/2\pi$ is
greater than two, $n>2$, a limit can be defined such that the condition for the
bulk theory to be free (\ref{free}) is satisfied. It turns out that for a
general irrational $n=2$ with a constant $c=1/\sqrt{5}$, and so in general a
free limit cannot be defined. Special irrationals with higher $n$ should allow a
free bulk. Does an M theory construction \cite{Witten} exist for them?

Let us {\it discuss the role of rationals} in these theories. It seems that there is
\underline{a continuous transition between theories of different
(rational) $\theta _6/2\pi$} thus answering a question raised in \cite{Witten}.
From a physical point of view note that the transition from rationals to
irrationals is smooth. From (\ref{qtension}) we see that as the rational
denominator, $q$, grows, the free bulk limit is reached later (for higher
$L_t$). Finally, when the number is irrational this limit cannot be reached
anymore. It is interesting to note another ``rational" physical phenomena - only
for rational $\theta _6/2\pi$ it is possible for a numbers of strips to join,
become a string and leave the 5-brane.

\begin{center}
\large{ACKNOWLEDGEMENTS}
\end{center}

The author thanks O. Aharony, E. Halyo, A. Rajaraman, L. Susskind and E. Witten.
BK is supported by NSF grant PHY-9219345.

\bibliography{irrat}
\bibliographystyle{utphys}

\end{document}